\documentclass[graybox]{svmult}

\usepackage{mathptmx}       
\usepackage{helvet}         
\usepackage{courier}        
\usepackage{type1cm}        
\usepackage{makeidx}         
\usepackage{graphicx}        
\usepackage{multicol}        
\usepackage[bottom]{footmisc}

\makeindex             

\begin{document}


\title*{Quantum queries associated with equi-partitioning of states and multipartite relational encoding across space-time}
\titlerunning{Equi-partitioning of quantum states and multipartite relational encoding}
\author{Karl Svozil}
\institute{Karl Svozil \at Institute for Theoretical Physics, Vienna
    University of Technology, Wiedner Hauptstra\ss e 8-10/136, A-1040
    Vienna, Austria, \email{svozil@tuwien.ac.at}}
%
%
\maketitle


\abstract{In the first part of this paper we analyze possible quantum computational capacities due to quantum queries associated with equi-partitions of pure orthogonal states. Special emphasis is given to the parity of product states and to functional parity. The second part is dedicated to a critical review of the relational encoding of multipartite states across (space-like separated) space-time regions; a property often referred to as ``quantum nonlocality.''}

\section{Unconventional properties for unconventional computing}
\label{sec:up}

At the heart of any unconventional form of computation (information processing) appears to be
some (subjectively and ``means relative'' to the current canon of knowledge)
strange, mind boggling, unexpected, stunning, surprising, hard to believe, feature or capacity of Nature.
That is,  in order to search for potentially unconventional information processing,
we have to parse for empirical patterns and behaviour as well as for theoretical predictions which,
relative to our expectations,  go beyond our everyday ``classical'' experience of the world.
``Unconventional'' always is ``means relative'' and has to be seen in a historic context;
 that is, relative to our present means and capacities which we consider consolidated and conventional.

For the sake of some example, take the transmission of data from one point to another via satellite links or cables;
or take (gps) navigation by time synchronization;
or take the prediction of all sorts of phenomena, including weather or astronomical events.
All these capacities appear conventional today, but would have been unconventional, or even magical, only 200 years ago.

So what are the new frontiers?
In what follows I shall mainly concentrate on some quantum physical capacities
which are widely considered as potential resources for presently ``unconventional'' computation.

Before we begin the discussion, a {\it caveat} is in order.
First, we should not get trapped by inappropriate yet convenient formal assumptions which have no operational consequences.
For instance, all kinds of ``infinity processes'' have no direct empirical correspondence.
In particular, classical continua abound in physics, but they need to be perceived rather
as convenient though metaphysical ``completion''
of processes and entities which are limited by finite physical means.

We should also not get trapped by what Jaynes called {\em Mind Projection Fallacy} \cite{jaynes-89,jaynes-90},
pointing out that
{\em ``we are all under an ego-driven temptation to project our private
thoughts out onto the real world, by supposing that the creations of one's own imagination are real
properties of Nature, or that one's own ignorance signifies some kind of indecision on the part of
Nature.''}
Instead we should attempt to maintain a curiosity with {\em evenly-suspended attention}
outlined by  Freud~\cite{Freud-1912} against {\em ``temptations to project,
what  [the analyst]  in dull self-perception recognizes as the peculiarities of his own personality,
as generally valid theory into science.''}

The postulate of ``true,'' that is, ontological, randomness in Nature is such a fallacy, in both
ways mentioned in the {\it caveat:} it assumes infinite physical resources (or maybe rather {\em ex nihilo} creation),
as well as our capacity to somehow being able to ``prove'' this --
a route blocked by recursion theory; in particular, by reduction to the Halting problem.

\section{Quantum speedups by equi-decomposition of sets of orthogonal states}
\label{2015-AUC-svozil-qsesos}

One of the mind boggling features of quantum information is that, unlike classical information,
it can ``reside,'' or be encoded into, the {\em relational properties} of multiple quanta~\cite{Zeilinger-97,zeil-99}.
For instance, the singlet Bell state of, say, two electrons is defined by the following property
(actually, two orthogonal spatial directions would suffice):
if one measures the spin properties of these particles along some arbitrary spatial direction,
then the spin value observed on one particle turns out to be always the negative spin value observed on the other particle
-- their relative spin value is negative -- that is,
either of ``$+$'' sign for the first particle and  of ``$-$'' sign  for the second one; or {\it vice versa}.
The ``rub,'' or rather compensation,  for this fascinating encoding of information ``across'' particles appears to be that
none of these individual particles has a definite individual spin value before the (joint) measurement.
That is, all information encodable into them is exhausted by these relational specifications.
This well recognized capacity of quantum mechanics
could be conceived as the ``essence of entanglement''~\cite{zeil-Zuk-bruk-01}.

Besides entanglement
there is another capacity which is not directly related to relational properties of multipartite systems;
yet it shares some similarities with the latter:
the possibility to organize elementary, that is, binary (or, in general,  $d$-ary)
quantum queries resolving properties which can be encoded into (equi-)partitions of some set of pure states.
If such partitions are feasible, then it is possible to obtain one bit (or, in general, dit) of information
by staging such a single query
{\em without} knowledge in what particular state the quantized system is.

From a different perspective any such (binary or $d$-ary) observable is
related to a {\em partial} (i.e. incomplete)
state identification~\cite{DonSvo01,DonSvo01vjqi,2007-tkadlec-svozil-springer}.
Many of the fast quantum algorithms discussed in the literature depend on incomplete state identification.

Note that, in the binary case, any complete state identification
--
that is, setting up a complete set of quantum observables or queries
capable to discriminate between and ``locating'' all single states
--
could be seen as the dual (observable) side of what can be considered an arbitrary state preparation
for multipartite systems.
This latter state preparation also features entanglement by allowing appropriate relational properties
among the constituent quanta.

\subsection{Parity of two-partite binary states}

For the sake of a demonstration of the ``unconventional'' quantum speedup achievable through partial (incomplete)
state identification, consider the four two-partite binary basis states
$\vert 00 \rangle$,
$\vert 01 \rangle$,
$\vert 10 \rangle$,
and
$\vert 11 \rangle$.
Suppose we are interested in the even parity of these states.
Then we could construct a {\em even parity operator}
$\textsf{\textbf{P}}$
{\em via} a spectral decomposition; that is,
\begin{equation}
\begin{array}{r}
\textsf{\textbf{P}} = 1 \cdot \textsf{\textbf{P}}_- + 0 \cdot \textsf{\textbf{P}}_+, \textrm { with}\\
\textsf{\textbf{P}}_- =  \vert 01 \rangle \langle 01 \vert + \vert 10 \rangle \langle 10 \vert,\\
\textsf{\textbf{P}}_+ =  \vert 00 \rangle \langle 00 \vert + \vert 11 \rangle \langle 11 \vert
,
\end{array}
\end{equation}
which yields even parity ``$0$'' on $\vert 00 \rangle$ as well as $\vert 11 \rangle$,
and
even parity ``$1$'' on $\vert 01 \rangle$ as well as $\vert 10 \rangle$, respectively.
Note that $\textsf{\textbf{P}}_-$
as well as
$\textsf{\textbf{P}}_+$
are projection operators, since they are idempotent; that is,
$\textsf{\textbf{P}}_-^2=\textsf{\textbf{P}}_-$ and
$\textsf{\textbf{P}}_+^2=\textsf{\textbf{P}}_+$.

Thereby, the basis of the two-partite binary states has been effectively equi-partitioned
into two groups of even parity ``$0$'' and ``$1$;'' that is,
\begin{equation}
\big\{
   \{\vert 00 \rangle ,  \vert 11 \rangle \},
   \{\vert 01 \rangle ,  \vert 10 \rangle \}
\big\}.
\end{equation}

The states associated with the propositions
corresponding to the projection operators
$\textsf{\textbf{P}}_-$ for even parity one
and
$\textsf{\textbf{P}}_+$ for even parity zero of the two bits
are entangled;
that is, this information is only expressed in terms of a {\em relational
property} -- in this case parity -- of the two quanta between each other~\cite{Zeilinger-97,zeil-99}.

\subsection{Parity of multi-partite binary states}

This equi-partinioning strategy~\cite{DonSvo01,svozil-2002-statepart-prl} to determine parity with a single query can be generalized
to determine the parity of multi-partite binary states.
Take, for example, the even parity of three-partite binary states definable by
\begin{equation}
\begin{array}{r}
\textsf{\textbf{P}} =  1 \cdot \textsf{\textbf{P}}_- + 0 \cdot \textsf{\textbf{P}}_+, \textrm { with}\\
\textsf{\textbf{P}}_- =
\vert 001 \rangle \langle 001 \vert  +
\vert 010 \rangle \langle 010 \vert  +
\vert 100 \rangle \langle 100 \vert  +
\vert 111 \rangle \langle 111 \vert,\\
\textsf{\textbf{P}}_+ =
\vert 000 \rangle \langle 000 \vert  +
\vert 011 \rangle \langle 011 \vert  +
\vert 101 \rangle \langle 101 \vert  +
\vert 110 \rangle \langle 110 \vert
.
\end{array}
\end{equation}

Again, the states associated with the propositions
corresponding to the projection operators
$\textsf{\textbf{P}}_-$ for even parity one
and
$\textsf{\textbf{P}}_+$ for even parity zero of the three bits
are entangled.
The basis of the three-partite binary states has been equi-partitioned
into two groups of even parity ``$0$'' and ``$1$;'' that is,
\begin{equation}
\begin{array}{r}
\big\{
   \{\vert 000 \rangle ,
\vert 011 \rangle ,
\vert 101 \rangle ,
\vert 110 \rangle  \}, \quad
\\
   \{\vert 001 \rangle ,
\vert 010 \rangle ,
\vert 100 \rangle ,
\vert 111 \rangle  \}
\big\}.
\end{array}
\end{equation}

\section{Parity of Boolean functions}

It is well known that Deutsch's problem -- to find out whether the output of a binary function of one bit
is constant or not; that is, whether the two outputs have even parity zero or one --
can be solved with one quantum query~\cite{nielsen-book,mermin-07}.
Therefore it might not appear totally unreasonable to
speculate that the parity of some Boolean function -- a binary function of an arbitrary number of bits --
can be determined by a single quantum query.
Even though we know that the answer is negative~\cite{Farhi-98}
it is interesting to analyze the reason why this parity problem is ``difficult''
even for quantum resources, in particular, quantum parallelism.
Because an answer to this question might provide us with insights about
the (in)capacities of quantum computations in general.

Suppose we define the functional parity $P(f_i)$ of an $n$-ary function $f_i = (g_i +1)/2$
{\it via} a function $g_i (  x_1,\ldots , x_n)\in \{-1,+1\}$ and
\begin{equation}
P(g_i) = \prod_{x_1,\ldots,x_n  \in \{0,1\}} g_i (  x_1,\ldots , x_n).
\end{equation}

Let us, for the sake of a direct approach of functional parity,
consider all the $2^{2^n}$ Boolean functions $f_i(  x_1,\ldots , x_n )$,
$0\le i \le 2^{2^n}-1$ of $n$ bits, and suppose that we can represent them by the
standard quantum oracle
\begin{equation}
\begin{array}{r}
U_i( \vert x_1,\ldots , x_n\rangle \vert y \rangle ) =\\
\vert x_1,\ldots , x_n\rangle \vert y\oplus f_i(   x_1,\ldots , x_n)\rangle
\end{array}
\end{equation}
as a means to cope with possible irreversibilities of the functions $f_i$.
Because $f_i\oplus f_i = 0$, we obtain $U_i^2 = {\bf I}$ and thus reversibility of the quantum oracle.
Note that all the resulting $n+1$-dimensional vectors are not necessarily mutually orthogonal.

For each particular $0\le i \le 2^{2^n}-1$, we can consider the set
\begin{equation}
F_i = \{ f_i (0,\ldots , 0),  \ldots , f_i (1,\ldots , 1) \}
\end{equation}
of all the values of $f_i$ as a function of all the $2^n$ arguments.
The set
\begin{equation}
\begin{array}{r}
V = \{F_i \mid 0\le i \le 2^{2^n}-1 \} \\
= \big\{ \{ f_i (0,\ldots , 0),  \ldots , f_i (1,\ldots , 1) \}\mid 0\le i \le 2^{2^n}-1 \big\}
\end{array}
\end{equation}
is formed by
all the $2^{2^n+n}$ Boolean functional values $f_i(  x_1,\ldots , x_n )$.
Moreover, for every one of the $2^{2^n}$ different Boolean functions of $n$ bits the $2^n$ functional output values
characterize the behavior of this function completely.

In the next step, suppose we equi-partition the set of all these functions into two groups:
those with even parity ``$0$'' and ``$1$,'' respectively.
The question now is this: can we somehow construct or find two mutually orthogonal subspaces (orthogonal projection operators)
such that all the parity ``$0$'' functions are represented in one subspace, and all the parity ``$1$'' are in the other, orthogonal one?
Because if this would be the case, then the corresponding (equi-)partition of basis vectors spanning those two subspaces
could be coded into a quantum query~\cite{DonSvo01} yielding the parity of $f_i$ in a single step.

We conjecture that involvement of one or more additional auxiliary bits
(e.g., to restore reversibility for nonreversible $f_i$'s)  cannot improve the situation,
as any uniform (over all the functions $f_i$) and
non-adaptive procedure
will not be able to generate proper orthogonality relations.

We know that for $n=1$ this task is feasible, since (we re-coded the functional value ``$0$'' to ``$-1$'')
\begin{equation}
\begin{array}{c|ccc}
f_i&P(f_i)&f_i(0)&f_i(1) \\
\hline
f_0& 0 & -1 & -1 \\
f_1& 0 & +1 & +1 \\
f_2& 1 & -1 & +1 \\
f_3& 1 & +1 & -1 \\
\end{array}
\end{equation}
and the two parity cases ``$0$'' and ``$1$,'' are coded into orthogonal subspaces spanned by $(1,1)$ and $(-1,1)$, respectively.

This is no longer true for $n=2$; due to an overabundance of functions,
the vectors corresponding to both parity cases ``$0$'' and ``$1$'' span the entire Hilbert space:
\begin{equation}
\begin{array}{c|ccccc}
f_i&P(f_i)&f_i(00)&f_i(01)&f_i(10)&f_i(11)\\
\hline
f_0& 0 & -1 & -1 & -1 & -1 \\
f_1& 0 & -1 & -1 & +1 & +1 \\
f_2& 0 & -1 & +1 & -1 & +1 \\
f_3& 0 & -1 & +1 & +1 & -1 \\
f_4& 0 & +1 & -1 & -1 & +1 \\
f_5& 0 & +1 & -1 & +1 & -1 \\
f_6& 0 & +1 & +1 & -1 & -1 \\
f_7& 0 & +1 & +1 & +1 & +1 \\
f_8& 1 & -1 & -1 & -1 & +1 \\
f_9& 1 & -1 & -1 & +1 & -1 \\
f_{10}& 1 & -1 & +1 & -1 & -1 \\
f_{11}& 1 & -1 & +1 & +1 & +1 \\
f_{12}& 1 & +1 & -1 & -1 & -1 \\
f_{13}& 1 & +1 & -1 & +1 & +1 \\
f_{14}& 1 & +1 & +1 & -1 & +1 \\
f_{15}& 1 & +1 & +1 & +1 & -1  \\
\end{array}
\end{equation}

\subsection{Proper specification of state discrimination}

The results of this section are also relevant for making precise Zeilinger's {\em foundational principle}~\cite{Zeilinger-97,zeil-99}
claiming that an $n$-partite system can be specified by exactly $n$ bits (dits in general).
The issue is what exactly is a ``specification?''

We propose to consider a specification appropriate if it can yield to an equi-partitioning of all pure states of the respective quantized system.
That is, to give an example, the parity of states could serve as a proper specification, but functional parity in general (for more that two quanta) cannot.

\section{Relativity theory {\it versus} quantum inseparability}
\label{sec:rt}

Let us turn our attention to another ``unconventional'' quantum resource,
which is mostly encountered at (but not restricted to) spatially separated entangled states: the so-called ``quantum nonlocality;''
and, in particular, on the paradigm shift of our perception of physical space and time.

First, let us keep in mind that, in the historic perspective it is quite evident why our current theory of space-time,
relativity theory~\cite{poincare02,ein-05},
does not directly refer to quanta: it was created
when quantum mechanics was ``unborn,'' or at least in its early infancy.
Indeed, in 1905 it was hardly foreseeable that Planck's self-denominated~\cite[p.~31]{hermann} {\it ``Akt der Verzweiflung''}
(``act of desperation'') -- committed five years ago in 1900 for the sake of
theoretically deriving precision measurements of the blackbody radiation --
would be extended into one of the most powerful physical theories imagined so far.
Therefore it should come as no surprise that all operationalizations and conventions implemented by relativity theory,
in particular, simultaneity, refer to classical, pre-quantum, physics.

Besides its applicability and stunning predictions and consequences (such as, for instance ``$E=mc^2$,''
as well as the unification of classical electric and magnetic phenomena)
the triumph of special relativity resides in its structural as well as formal clarity:
by adopting certain conventions
(which were essentially adopted from railroad traffic~\cite{Galison-2000,Galison-2003}
and are also used by {\em Cristian's Algorithm} for data network synchronization),
and by fixing the speed of electromagnetic radiation for all reference frames
(together with the requirement of bijectivity),
the Lorentz transformations result from theorems of incidence geometry~\cite{lester,naber}.
Beyond formal conventions, the physical content resides in the form invariance of the equations of motion
under such transformations.

In view of these sweeping successes of classical relativity theory it might not be surprising that Einstein,
one of the creators of quantum mechanics,
never seriously considered the necessity to adapt the concepts of space-time to the new quantum physics.
On the contrary -- Einstein seemed to have prioritized relativity over quantum theory;
the latter one he critically referred to as~\cite[p.~113]{born-69} {\it ``noch nicht der wahre Jakob''
(``not yet the true [final] answer'').}
Time and again Einstein
came up with predictions of quantum mechanics which allegedly discredited the (final) validity of quantum theory.

In a letter  to Schr\"odinger dated June 19th, 1935~\cite{einstei-letter-to-schr,Howard1985171}
Einstein concretized and clarified
his uneasiness with quantum theory
previously published in a paper with Podolsky and Rosen~\cite{epr} ({\it ``written by Podolsky
after many discussions''}~\cite{einstei-letter-to-schr}).
In this communication
Einstein insisted that
the wave function of a subsystem $A$ of (entangled) particles cannot
depend on whatever measurements are performed on its spatially
separated (i.e. separated by a space-like interval) ``twin'' subsystem $B$:
in his own (translated from German~\footnote{Einstein's (underlined) original German text:
{\it ``Der wirkliche Zustand von $B$ kann nicht davon abh\"angen,
was f\"ur eine Messung ich an $A$ vornehme.''}}) words:
{\it ``The true state of $B$ cannot depend on what measurement I perform on $A$.''}
Pointedly stated, the  {\it ``separability principle'' asserts
that any two spatially separated systems possess their own separate real state}~\cite{Howard1985171}.

The separability principle is not satisfied for entangled states~\cite{CambridgeJournals:1737068,schrodinger};
in particular, if general two-partite state
$$\vert \Psi \rangle
=
\sum_{i,j \in \{-,+\} }
\alpha_{ij} \vert ij \rangle
\textrm{,
with
}
\sum_{i,j \in \{-,+\} }
\vert \alpha_{ij} \vert^2  =1
$$
does not satisfy factorizability~\cite[p.~18]{mermin-07} requiring
$\alpha_{--}\alpha_{++} = \alpha_{+-}\alpha_{-+}$.
That is, if $\alpha_{--}\alpha_{++} \neq \alpha_{+-}\alpha_{-+}$,
then $\vert \Psi \rangle  $ cannot be factored into products of single particle states.

Even in his later years Einstein was inclined to take relativistic space-time as the primary framework;
thereby prioritizing it
over fundamental quantum mechanical inseparability; in particular,
when it comes to multipartite situations~\cite{Einstein-48,Howard1985171}.

\section{Proximity and apartness in quantum mechanics}
\label{sec:2}

In what follows we propose that,
when it comes to microphysical situations,
in particular, when entanglement is involved,
the provenance of classical relativity theory over quantum mechanics
has to be turned upside down:
while entangled quanta may epistemically (and for many practical purposes~\cite{bell-a})
appear ``separated,'' or ``apart,'' or
``distinct''
to a classical observer ignorant of their relational properties
(cf. earlier discussion in Section \ref{2015-AUC-svozil-qsesos})
encoded ``across these quanta,''
quantum mechanically they are treated holistically ``as one.''

The pretension of any such observer, or the possibility to actually perceive
entangled quanta as being ``spatially separated'' (by disregarding their correlations)
should not be seen as a principal property, but rather as a {\em ``means relative''} one.

For the sake of an example, take the two-particle singlet Bell state
$\vert \Psi^- \rangle
= (1/\sqrt{2})\left(\vert +- \rangle - \vert -+ \rangle \right)$,
which, by identifying  $\vert - \rangle \equiv (0,1)$ and $\vert + \rangle \equiv (1,0)$,
can be identified with the four-dimensional vector
whose components in tuple form are $\vert \Psi^- \rangle  \equiv
(1/\sqrt{2})[(1,0)\otimes (0,1)  - (0,1) \otimes (1,0)]
=
(1/\sqrt{2})(0,1,-1,0)$.
The separability principle is not satisfied, since $0\cdot 0 \neq 1 \cdot (-1)$.
So, from the point of view of those entangled state observables,
the quanta appear inseparable.

And yet, the same quanta can be perfectly localized and distinguished by
resolving them spatially.
This situation -- the occurrence of both inseparability and (spatial) distinguishability --
has caused a lot of confusion.
This is particularly serious if one of these distinct viewpoints on the quantized system,
say, spatial separability and locatedness of the particles,
is meshed
with the inseparability of the spin observables when the latter ones are relationally defined.
An yet, we might envision that, with this dual situation we could get a handle
on quantum inseparability ({\it via} encoding of relational information)
by spatially separated detectability of the quanta forming this entangled state.
Alas this is impossible, because the relational properties do not reveal themselves by individual outcomes --
only when all these (relational) outcomes are considered together do the relational properties reveal themselves.

Of course, this would be totally different if it would be possible to willfully {\em force}
any particular handle or side or component of the entangled state, thereby effectively
forcing the respective (relational property on the other handle or side or component.
So far, despite speculative attempts to utilize stimulated emission~\cite{svozil-slash},
there is no indication that this might be physically feasible.

\section{Summary}

The first part of this article has been dedicated to quantum queries relating to properties
which can be encoded in terms of (equi-)partitioning of states.
We have been particularly interested in the parity of products of binary states, and also
in the parity of Boolean functions; that is, dichotomic functions of bits.
Thereby we have presented criteria for the (non-) existence of quantum oracles.

In the second part of this article we have argued that,
instead of perceiving entangled quanta in an {\it a priori} ``space-time theater,''
space-time is a secondary, derived concept of our mind which needs to be operationally constructed by
conventions and observations. This is particularly true for multipartite entangled states,
and their spatio-temporal interconnectedness.
Such an approach leaves no room for any hypothetical inconsistency in quantum space-time,
and no mind-boggling ``peaceful coexistence'' with relativity theory.

\begin{acknowledgement}
This research has been partly supported by FP7-PEOPLE-2010-IRSES-269151-RANPHYS.
\end{acknowledgement}


\begin{thebibliography}{10}

\bibitem{jaynes-89}
Edwin~Thompson Jaynes.
\newblock Clearing up mysteries - the original goal.
\newblock In John Skilling, editor, {\em Maximum-Entropy and Bayesian Methods:
  : Proceedings of the 8th Maximum Entropy Workshop, held on August 1-5, 1988,
  in St. John's College, Cambridge, England}, pages 1--28. Kluwer, Dordrecht,
  1989.

\bibitem{jaynes-90}
Edwin~Thompson Jaynes.
\newblock Probability in quantum theory.
\newblock In Wojciech~Hubert Zurek, editor, {\em Complexity, Entropy, and the
  Physics of Information: Proceedings of the 1988 Workshop on Complexity,
  Entropy, and the Physics of Information, held May - June, 1989, in Santa Fe,
  New Mexico}, pages 381--404. Addison-Wesley, Reading, MA, 1990.

\bibitem{Freud-1912}
Sigmund Freud.
\newblock {R}atschl{\"{a}}ge f{\"{u}}r den {A}rzt bei der psychoanalytischen
  {B}ehandlung.
\newblock In Anna Freud, E.~Bibring, W.~Hoffer, E.~Kris, and O.~Isakower,
  editors, {\em {G}esammelte {W}erke. {C}hronologisch geordnet. {A}chter
  {B}and. {W}erke aus den {J}ahren 1909--1913}, pages 376--387, Frankfurt am
  Main, 1999. Fischer.

\bibitem{Zeilinger-97}
Anton Zeilinger.
\newblock Quantum teleportation and the non-locality of information.
\newblock {\em Philosophical Transactions of the Royal Society of London A},
  355:2401--2404, 1997.

\bibitem{zeil-99}
Anton Zeilinger.
\newblock A foundational principle for quantum mechanics.
\newblock {\em Foundations of Physics}, 29(4):631--643, 1999.

\bibitem{zeil-Zuk-bruk-01}
{\v{C}}aslav Brukner, Marek Zukowski, and Anton Zeilinger.
\newblock The essence of entanglement.
\newblock Translated to Chinese by Qiang Zhang and Yond-de Zhang, New Advances
  in Physics (Journal of the Chinese Physical Society), 2002.

\bibitem{DonSvo01}
Niko Donath and Karl Svozil.
\newblock Finding a state among a complete set of orthogonal ones.
\newblock {\em Physical Review A}, 65:044302, 2002.

\bibitem{DonSvo01vjqi}
Niko Donath and Karl Svozil.
\newblock Finding a state among a complete set of orthogonal ones.
\newblock {\em Virtual Journal of Quantum Information}, 2, April 2002.

\bibitem{2007-tkadlec-svozil-springer}
Karl Svozil and Josef Tkadlec.
\newblock On the solution of trivalent decision problems by quantum state
  identification.
\newblock {\em Natural Computing}, in print:in print, 2009.

\bibitem{svozil-2002-statepart-prl}
Karl Svozil.
\newblock Quantum information in base $n$ defined by state partitions.
\newblock {\em Physical Review A}, 66:044306, 2002.

\bibitem{nielsen-book}
M.~A. Nielsen and I.~L. Chuang.
\newblock {\em Quantum Computation and Quantum Information}.
\newblock Cambridge University Press, Cambridge, 2000.

\bibitem{mermin-07}
David~N. Mermin.
\newblock {\em Quantum Computer Science}.
\newblock Cambridge University Press, Cambridge, 2007.

\bibitem{Farhi-98}
Edward Farhi, Jeffrey Goldstone, Sam Gutmann, and Michael Sipser.
\newblock Limit on the speed of quantum computation in determining parity.
\newblock {\em Physical Review Letters}, 81:5442--5444, 1998.

\bibitem{poincare02}
Henri Poincar{\'{e}}.
\newblock {\em La Science et l'hypoth\'ese}.
\newblock Flammarion, Paris, 1902.

\bibitem{ein-05}
Albert Einstein.
\newblock {Z}ur {E}lektrodynamik bewegter {K}{\"{o}}rper.
\newblock {\em Annalen der Physik}, 322:891--921, 1905.

\bibitem{hermann}
Armin Hermann.
\newblock {\em {F}r\"uhgeschichte der {Q}uantentheorie (1899-1913)}.
\newblock {P}hysik {V}erlag, Mosbach in Baden, 1969.
\newblock {H}abilitationsschrift, Naturwissenschaftliche Fakult\"at der
  Universit\"at M\"unchen.

\bibitem{Galison-2000}
Peter~Louis Galison.
\newblock {E}instein's clocks: The place of time.
\newblock {\em Critical Inquiry}, 26(2):355--389, 2000.

\bibitem{Galison-2003}
Peter~Louis Galison.
\newblock {\em {E}instein's clocks, {P}oincar\'e's maps: Empires of Time}.
\newblock W.W. Norton \& Company, New York and London, 2003.

\bibitem{lester}
June~A. Lester.
\newblock Distance preserving transformations.
\newblock In Francis Buekenhout, editor, {\em Handbook of Incidence Geometry},
  pages 921--944. Elsevier, Amsterdam, 1995.

\bibitem{naber}
Gregory~L. Naber.
\newblock {\em The Geometry of {M}inkowski Spacetime}, volume~92 of {\em
  Applied Mathematical Sciences}.
\newblock {ANU} {Q}uantum {O}ptics, New York, Dordrecht, Heidelberg, London,
  second edition, 2012.

\bibitem{born-69}
Max Born.
\newblock {\em Physics in my generation}.
\newblock Springer, New York, second edition, 1969.

\bibitem{einstei-letter-to-schr}
Albert Einstein.
\newblock Letter to {S}chr\"odinger.
\newblock Old Lyme, dated 19.6.35, Einstein Archives 22–047 (searchable by
  document nr. 22-47), 1935.

\bibitem{Howard1985171}
Don Howard.
\newblock {E}instein on locality and separability.
\newblock {\em Studies in History and Philosophy of Science Part A},
  16(3):171--201, 1985.

\bibitem{epr}
Albert Einstein, Boris Podolsky, and Nathan Rosen.
\newblock Can quantum-mechanical description of physical reality be considered
  complete?
\newblock {\em Physical Review}, 47(10):777--780, May 1935.

\bibitem{CambridgeJournals:1737068}
Erwin Schr{\"{o}}dinger.
\newblock Discussion of probability relations between separated systems.
\newblock {\em Mathematical Proceedings of the Cambridge Philosophical
  Society}, 31(04):555--563, 1935.

\bibitem{schrodinger}
Erwin Schr{\"{o}}dinger.
\newblock Die gegenw{\"{a}}rtige {S}ituation in der {Q}uantenmechanik.
\newblock {\em Naturwissenschaften}, 23:807--812, 823--828, 844--849, 1935.

\bibitem{Einstein-48}
A.~Einstein.
\newblock {Q}uanten-{M}echanik und {W}irklichkeit.
\newblock {\em Dialectica}, 2(3-4):320324, 1948.

\bibitem{bell-a}
John~S. Bell.
\newblock Against `measurement'.
\newblock {\em Physics World}, 3:33--41, 1990.

\bibitem{svozil-slash}
Karl Svozil.
\newblock What is wrong with {SLASH}?
\newblock eprint arXiv:quant-ph/0103166, 1989.

\end{thebibliography}

\printindex
\end{document}